\documentclass[aps,prl,twocolumn,groupedaddress,longbibliography]{revtex4-1}
\usepackage[dvipsnames]{xcolor}
\usepackage{graphicx,amsmath,amssymb,color}
\usepackage{soul}

\newcommand{\changedforRone}[1]{{#1}}
\newcommand{\changedforRtwo}[1]{{#1}}
\newcommand{\otherchanges}[1]{{#1}}

\begin{document}

\title{Critical\changedforRone{-like slowdown} in thermal soft-sphere glasses via energy minimization}
\author{Kevin A. Interiano-Alberto$^1$, Peter K. Morse$^2$, and Robert S. Hoy$^{1}$}
\email{rshoy@usf.edu}
\affiliation{$^1$Department of Physics, University of South Florida, Tampa, Florida 33620,USA}
\affiliation{$^2$Department of Chemistry, Department of Physics, and Princeton Institute of Materials, Princeton University, Princeton, New Jersey 08544, USA}
\date{\today}
\begin{abstract}
Using hybrid molecular dynamics/SWAP Monte Carlo (MD/SMC) simulations, we show that the terminal relaxation times $\tau$ for FIRE energy minimization of soft-sphere glasses exhibit thermal onset as samples become increasingly well-equilibrated. 
\otherchanges{Although} $\tau(\phi)$ can decrease by orders of magnitude as equilibration proceeds and the jamming density $\phi_{\rm J}$ increases via thermal onset, it always scales as $\tau(\phi) \sim (\phi_{\rm J} - \phi)^{-2} \sim [Z_{\rm iso} - Z_{\rm ms}(\tau)]^{-2}$, where $\phi_{\rm J}$ is the jamming density and $Z_{\rm ms}(\tau)$ is the average coordination number of particles satisfying a minimal local mechanical stability criterion ($Z \geq d+1$) \textit{at the top of} the final \otherchanges{potential-energy-landscape (PEL)} sub-basin the system encounters.
This scaling allows us to collapse $\tau$ datasets that look very different when plotted as a function of $\phi$, and to address a closely related question: \changedforRtwo{how does the character of the PEL basins that dense thermal glasses \textit{most typically} occupy evolve as the glasses age at constant $\phi$ and $T$?}
\end{abstract}
\maketitle

Jamming exhibits many features that are \otherchanges{reminscent of} critical phenomena \cite{ohern03}.
\otherchanges{Since multiple} length and time scales exhibit power-law divergences as $\phi_{\rm J}$ is approached from below \cite{wyart05,keys07,hopkins12a,berthier16b}, \otherchanges{so do their} associated mechanical quantities. 
For example, the shear viscosity of colloidal suspensions ($\eta$), which is often assumed to be linearly proportional to their characteristic stress-relaxation time $\tau_{\rm visc}$, scales as \changedforRone{$\eta \sim |\Delta\phi|^{-\beta}$, where $\Delta\phi = \phi-\phi_{\rm J}$ is the excess packing fraction and $1.6 \leq \beta \leq 4$ \cite{brady93,olsson07,forterre08,nordstrom10,boyer11,andreotti12,kawasaki15,suzuki15}.
Correspondingly, recent simulations have shown that the characteristic relaxation times ($\tau^*$) for energy minimization and shear-stress relaxation in athermal hard and soft sphere glasses scale as  $\tau^* \sim |\Delta Z|^{-\nu}$, where $\Delta Z = Z - Z_{\rm iso} \equiv Z - 2d$ is the excess coordination number, and $1.6 \lesssim \nu \lesssim 3.7$ \cite{heussinger09,olsson19,lerner12,lerner12b,olsson15,ikeda20,saitoh20,nishikawa21}.
These divergences} can be understood in terms of the relation  $\tau^* \sim \omega^{-2}_{\rm min}$, where $\omega_{\rm min}$ is the frequency of systems' lowest-energy vibrational mode \cite{lerner12,lerner12b}.
Such modes increasingly dominate  systems'  \otherchanges{relaxation dynamics} as $\phi \to \phi_{\rm J}$ from below and $\omega_{\rm min} \to 0$  \cite{lerner12,lerner12b,olsson15,ikeda20,saitoh20,footcalcDOS2}.
Assuming that they control $\eta$ for densities just below jamming, and employing the relation $\Delta Z \sim \Delta \phi$,  
allows the  abovementioned scaling relation to be re-expressed as $\eta \sim \tau^* \sim |\Delta Z|^{-\nu}$, if in fact $\beta = \nu$ \otherchanges{as suggested by Refs.\ \cite{heussinger09,lerner12,lerner12b,olsson15,ikeda20,saitoh20,olsson19}.}

A recent study \cite{nishikawa21} has challenged some of the main conclusions of Refs.\ \cite{lerner12,lerner12b,olsson15,ikeda20,saitoh20}, and in particular their  assertions that (i) the divergence of $\tau^*$ represents a true critical phenomenon with a well-defined value of $\nu$, and (ii) athermal glasses' $\tau_{\rm visc}$ and $\eta$ are \otherchanges{both} controlled by $\tau^*$.
\otherchanges{On the other hand,} the critical-\changedforRone{\textit{like} slowdown} of athermal soft-sphere glasses' energy-minimization and shear-stress-relaxation dynamics as $\phi \to \phi_{\rm J}$ and $Z \to Z_{\rm iso}$ from below is now well-established.
The extent to which these phenomena affect \textit{thermal} glasses' relaxation dynamics, and hence are subject to ``onset'' effects, however, has not been explored.
\changedforRone{Thermalized} 3D hard-sphere liquids equilibrated at packing fractions $\phi_{\rm eq}$ below the onset density \changedforRone{$\phi_{\rm on} \simeq 0.45$ always} have the same jamming density $\phi_{\rm J} = \phi_{\rm RCP} \simeq 0.64$ \cite{torquato00}, while those equilibrated at $\phi_{\rm \otherchanges{eq}} > \phi_{\rm on}$ have $\phi_{\rm J}$ that increase with increasing $\phi_{\rm eq}$, or, for fixed $\phi_{\rm eq}$, with increasing equilibration time $t_{\rm eq}$ \cite{torquato00,chaudhuri10,ozawa17,morse21}.
Similarly, soft-sphere liquids  equilibrated  at \changedforRone{fixed $\phi$ and} temperatures $T$ above the onset temperature \changedforRone{$T_{\rm on}(\phi)$ always} have the same average inherent structure energy ($E_{\rm IS}$), while those equilibrated at temperatures $T <  T_{\rm on}$ have $E_{\rm IS}$ that decrease with decreasing $T$ and increasing $t_{\rm eq}$ \cite{sastry98,debenedetti01}.
Because Refs.\  \cite{brady93,olsson07,forterre08,heussinger09,nordstrom10,boyer11,andreotti12,kawasaki15,suzuki15,lerner12,lerner12b,olsson15,ikeda20,saitoh20,olsson19,nishikawa21} all examined systems where $\phi_{\rm J} \simeq \phi_{\rm RCP}$, a natural followup question is: how are the divergences of time scales like $\tau^*$  affected by sample preparation\changedforRone{/thermal onset, i.e.\ by the abovementioned increasing $\phi_{\rm J}(t_{\rm eq})$ and decreasing $T_{\rm on}(\phi,t_{\rm eq})$?}

In this Letter, using  MD/SMC simulations combined with \otherchanges{FIRE} energy minimization \cite{grigera01,ninarello17,bitzek06,guenole20},  we shed light on this question.
By starting with far-from-equilibrium soft-sphere glasses obtained via infinite-temperature quenches (with a wide range of $\phi$) and then bringing them towards equilibrium using SWAP, we show that the  the times $\tau$ required for \textit{thermal} soft-sphere glasses to enter their final unjammed potential-energy-landscape (PEL) sub-basin during energy minimization exhibit thermal onset as samples become increasingly well-equilibrated, in the same fashion that $\phi_{\rm J}(t_{\rm eq})$ does.
Although  \changedforRone{$\tau(\phi,t_{\rm eq})$} can decrease by orders of magnitude as equilibration proceeds and $\phi_{\rm J}(t_{\rm eq})$ increases via thermal onset, it always scales as \changedforRone{$\tau(\phi,t_{\rm eq}) \sim [\phi_{\rm J}(t_{\rm eq}) - \phi]^{-2} \sim \Delta Z^{-2}$}, where \changedforRone{$\Delta Z \equiv Z_{\rm iso} - Z_{\rm ms}[\tau(\phi,t_{\rm eq})]$}, for sufficiently small $\Delta \phi$ and $\Delta Z$.
\changedforRtwo{This common scaling allows us to collapse $\tau$ datasets that look very different when plotted as a function of $\phi$ and $t_{\rm eq}$, and thus to greatly simplify our understanding of dense thermal soft-sphere glasses' strongly $\phi$- and $t_{\rm eq}$-dependent energy-minimization dynamics.}

All simulations were performed using hdMD \cite{hoy22}. 
Systems are initialized by placing $N = 10^5$ soft-sphere particles randomly within periodic 3D cubic simulation cells, with a wide range of packing fractions ($0.63 \leq \phi \leq 0.68$).
\otherchanges{These particles are continuously-}polydisperse, with a size distribution that \otherchanges{produces} excellent glass-forma\otherchanges{bility} for a variety of pair potentials \changedforRtwo{\cite{ninarello17,SuppMat,footd23}}.
Infinite-temperature quenches are performed \otherchanges{\cite{ohern03}, and then systems} are equilibrated at \otherchanges{a constant temperature} $k_B T_{\rm eq} = \tilde{\epsilon}$ using the SWAP algorithm \cite{grigera01,ninarello17}, for times $t_{\rm eq}$ up to $10^5\tilde{\tau}$.
\changedforRtwo{Our implementation attempts $N/10$ particle-diameter swaps per $\tilde{\tau}$.} 
Here $\tilde{\tau} = \sqrt{\tilde{m}\tilde{\sigma}^2/\tilde{\epsilon}}$ is the unit of time and $\tilde{m}$, $\tilde{\sigma}$, and $\tilde{\epsilon}$ are respectively the units of mass, length, and energy\otherchanges{; below, we will express all quantities in terms of these units.}
For most $\phi$ examined here, this procedure produces weakly-to-moderately-aged glasses (i.e.\ \textit{not} equilibrated supercooled liquids), consistent with our goal of studying nonequlibrium phenomena that occur deep in the glassy state.

At selected \otherchanges{values of} $t_{\rm eq}$, we minimize systems' energies using the FIRE  \cite{bitzek06,guenole20} algorithm.
During these minimizations, we monitor changes in the average pair energy per particle $E_{\rm p} = N^{-1} \sum_{j > i} U(r_{ij})$ as well as $Z$ and $Z_{\rm ms}$, which are respectively the average coordination numbers for all particles and for \changedforRtwo{all particles $i$ that satisfy a \textit{minimal, local} mechanical stability criterion $Z_i \equiv \sum_{j \neq i} \Theta(\sigma_{ij} - r_{ij}) \geq d+1$ \cite{footZdp1,morse23}.
Here $r_{ij}$ is the distance between particles $i$ and $j$, $\sigma_{ij}$ is their reduced interparticle diameter \cite{SuppMat}, $\Theta$ is the Heaviside step function, and interparticle contacts are identified using the standard criterion $r_{ij} < \sigma_{ij}$ \cite{ohern03}.}
Below, we plot these quantities as a function of the elapsed minimization time 
\begin{equation}
t = \displaystyle\sum_{i = 0}^I \delta t_i
\label{eq:tdef}
\end{equation}
after $I$ FIRE iterations, where $\delta t_i$ is the adaptive timestep during the $i$th iteration \cite{guenole20}.
Energy minimization continues until $E_{\rm p}$ reaches $0$, $E_{\rm p}$ has not changed over the past ten iterations, or $I$ reaches $10^5$ \cite{SuppMat}. 
\changedforRone{Since FIRE dynamics are only partially physical (in contrast to steepest-descent dynamics, which correspond to the limit of infinite damping \cite{SuppMat,foot2p7,footsmalltauF}), we will not assign any physical significance to \textit{absolute} values of $t$; below, we will only make \textit{relative} statements.}

\begin{figure}[htbp]
\includegraphics[width=3.125in]{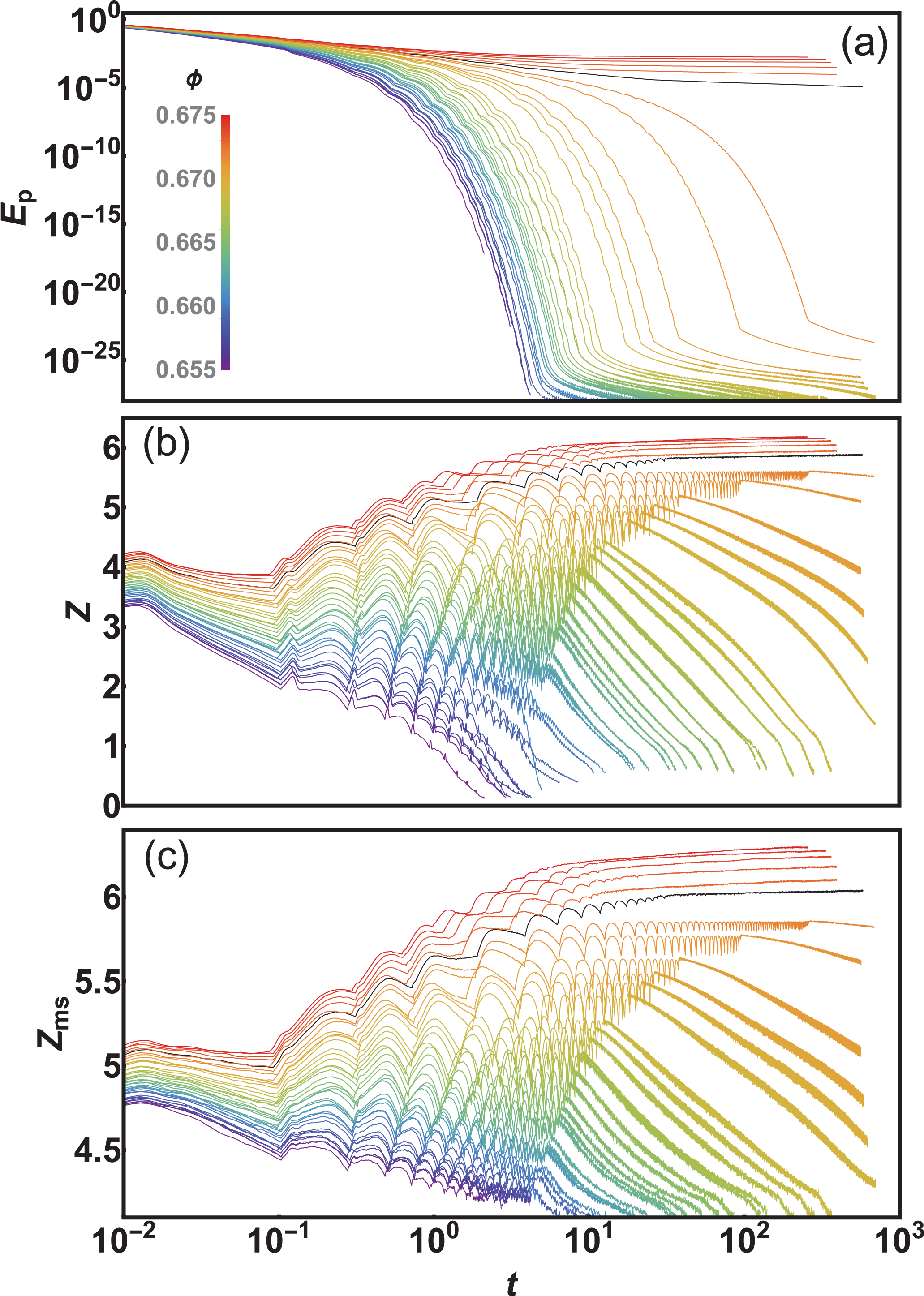}
\caption{Structural metrics during FIRE energy minimization of 3D thermal soft-sphere glasses equilibrated for $t_{\rm eq} = 6\times 10^4$.  
This system has $\phi_{\rm J}(t_{\rm eq}) = 0.6726$ (from Eq.\ \ref{eq:tkphi})\changedforRtwo{; black curves indicate results for the lowest $\phi \gtrsim \phi_{\rm J}$.}}
\label{fig:firesummary}
\end{figure}

\otherchanges{We begin by discussing the $\phi$- and $t$-dependence of $E_{p}$, $Z$, and $Z_{ms}$ for one representative $t_{\rm eq}$ value ($6\times 10^4$).}
Figure \ref{fig:firesummary}(a) shows $E_{\rm p}(t)$ data for systems with $.655 \leq \phi \leq .675$, in increments $\delta \phi = .0005$.
During the initial stages of energy minimization, all systems have $E_{\rm p}(t) \sim t^{-1}$ \cite{ikeda20}. 
In jammed systems,  $\partial E_{\rm p}/\partial t$ increases monotonically with $t$, and \otherchanges{$E_{\rm p}$} converges faster as $\phi$ increases, consistent with previous studies \cite{olsson22,nishikawa22,manacorda22,charbonneau23}.
For unjammed systems, the response is qualitatively different.
\otherchanges{The initial $E_{\rm p}(t) \sim t^{-1}$ regimes end at times $t_{\rm drop}(\phi)$.}
Over a wide range of $\phi$ and $t > t_{\rm drop}\otherchanges{(\phi)}$,  $E_{\rm p}(t) \sim \exp[-t/\tau^*(\phi)]$, where (consistent with previous studies \cite{lerner12,lerner12b,olsson15,ikeda20,saitoh20}) $\tau^* \sim (\phi_{\rm J}-\phi)^{-2}$ as $\phi \to \phi_{\rm J}$ from below \otherchanges{\cite{SuppMat}}.
However, $E_{\rm p}$ does not smoothly drop all the way to zero as might be expected.
Instead, the \otherchanges{exponentially-decaying} portions of the $E_{\rm p}(t)$ curves end at finite $E_{\rm p}$, exhibiting kinks at times $ \tau(\phi)$ that increase rapidly with $\phi$.
During the final stages of minimization, $E_{\rm p}$ drops towards zero in a roughly power-law fashion.
Overall, the $E_{\rm p}(t)$ dataset suggests that the kinks for $\phi < \phi_{\rm J}$ correspond to systems entering their final PEL sub-basin.

This hypothesis is strongly supported by examining the coordination number $Z(t)$.
As shown in Fig.\ \ref{fig:firesummary}(b), the $Z(t)$ exhibit a common behavior for \changedforRtwo{small $t$, first decreasing and then increasing} as the elimination of strong interparticle overlaps brings more particles into contact with each other.
For $\phi \geq \phi_{\rm J}$, these increases persist to $t \to \infty$ as is typical of jammed systems \cite{ohern03}.
For $\phi < \phi_{\rm J}$, however, they terminate at the same finite $ \tau(\phi)$ shown in panel (a).
For $t >  \tau(\phi)$, the $Z(t)$ [much like the $E_{\rm p}(t)$] drop slowly towards zero.
At intermediate times, $Z(t)$ oscillates.
The local minima in $Z(t)$ coincide with the FIRE algorithm resetting when the system encounters a saddle point and the dot product of the $N$-particle force and velocity vectors ($\vec{F}\cdot\vec{v}$) for a \textit{prospective} set of particle positions $\{ r \}$ becomes negative \cite{bitzek06}.
After these resetting events, $Z$ tends to first increase as a few larger interparticle overlaps get converted into many smaller ones, then decrease again as these small overlaps are eliminated.
Since this occurs when the system traverses a region in which the direction of $\vec{F}$ is changing substantially from one iteration to the next, the oscillations cease once it has entered its final PEL sub-basin [at  $t =  \tau(\phi)$].
Fig.\ \ref{fig:firesummary}(c) shows that the character of these oscillations is not changed by removing particles with $\changedforRtwo{Z_i} < d+1$ \cite{footZdp1,SuppMat}.
\changedforRtwo{Note, however, that for both $Z(t)$ and $Z_{\rm ms}(t)$ their amplitude decreases and their frequency increases as $\phi \to \phi_{\rm J}$.}

We find that $\tau$ is always linearly proportional to (albeit substantially larger than \cite{SuppMat}) $\tau^*$, indicating that the results reported above are closely related to those discussed in Refs.\  \cite{lerner12,lerner12b,olsson15,ikeda20,saitoh20}.
Since these studies employed either normal MD time integration (in simulations of shear stress relaxation) or gradient-descent rather than FIRE energy minimization, they were unable to observe the kinks in $E_{\rm p}(t)$ and oscillations in $Z(t)$ and $Z_{\rm ms}(t)$ discussed above, or \otherchanges{to} measure an exact analogue to the terminal relaxation time $ \tau$.
As we will demonstrate below, the utility of the above discussion is that it allows us to convincingly argue that $\Delta Z(\tau) \equiv Z_{\rm iso} - Z_{\rm ms}( \tau)$, i.e.\  
minimally-locally-stable particles' average hypostaticity \textit{at the top of} the final sub-basin the system encounters, is a well-defined quantity that can be used to describe these systems' energy-minimization dynamics.

\begin{figure}[h]
\includegraphics[width=3.125in]{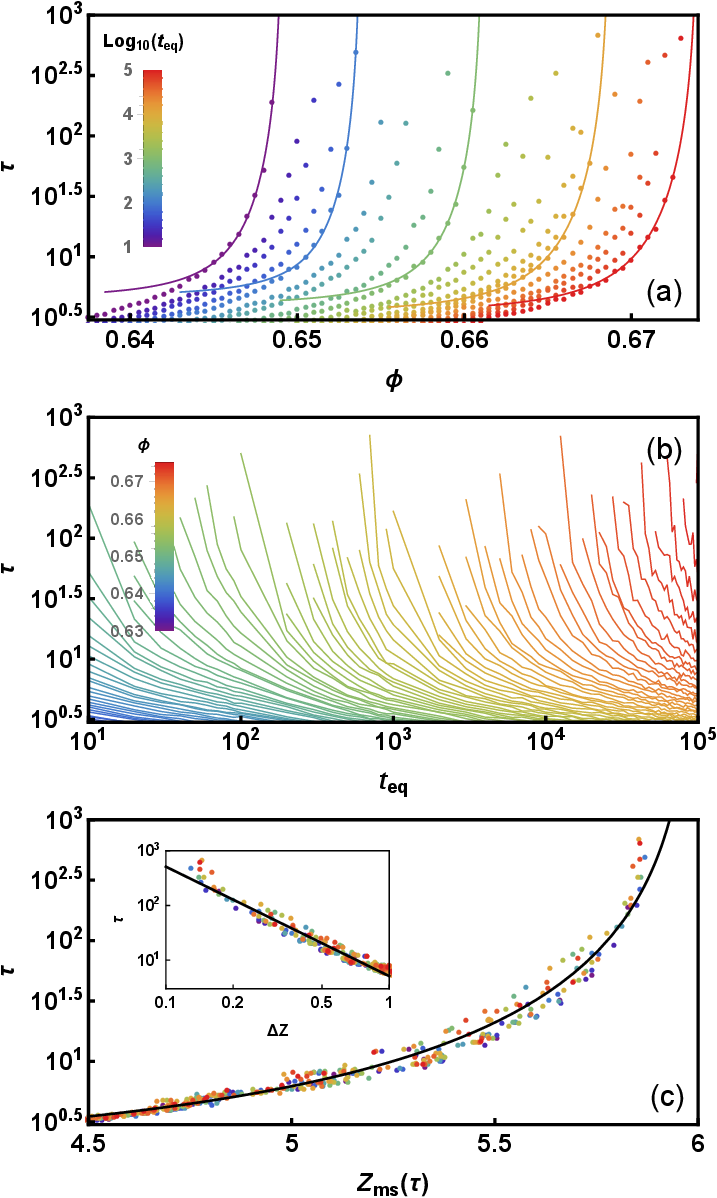}
\caption{Effect of thermal onset on soft-sphere glasses' energy-minimization dynamics.  
Panels (a) and (c) respectively show  $ \tau$ vs.\ $\phi$ and $Z_{\rm ms}( \tau)$ for selected $t_{\rm eq}$, while panel (b) shows \otherchanges{$\tau$ vs.\ $t_{\rm eq}$} for selected $\phi$.  Solid curves in panel (a) show fits to Eq.\ \ref{eq:tkphi} for $t_{\rm eq} = 10^1, 10^2, 10^3, 10^4,\ \! \rm{and}\ 10^5$.
In panel (c), the color legend in the same as in panel (a), the solid curve  shows a single fit of the entire dataset to Eq.\ \ref{eq:tkZnr} with $C = 1.55$ and $D = 4.86$, and the inset shows the same data plotted vs.\ $\Delta Z$, with a line indicating  $\Delta Z^{-2}$ scaling.}
\label{fig:fire2}
\end{figure}

Our main contribution centers around the fact that the only substantial changes in the phenomona illustrated in Fig.\ \ref{fig:firesummary} as $t_{\rm eq}$ increases are that they shift to higher $\phi$, following the increase in $\phi_{\rm J}(t_{\rm eq})$ as thermal onset proceeds.
\changedforRone{We observe exponential decay of $E_p$ terminating in kinks at $t = \tau(\phi,t_{\rm eq})$, oscillations in $Z$ and $Z_{\rm ms}$, and divergences in $\tau \sim \tau^*$ as $\phi \to \phi_{\rm J}(t_{\rm eq})$ from below, for all values of $t_{\rm eq}$ \cite{SuppMat}.}
Results for \changedforRone{$\tau(\phi,t_{\rm eq})$} for a wide range of $\phi$ and $t_{\rm eq}$ are summarized in Figure \ref{fig:fire2}.
Panel (a) shows how the jamming densities $\phi_{\rm J}(t_{eq})$ obtained by fitting the finite [$\phi < \phi_{\rm J}(t_{\rm eq})$] $ \tau$ values to the empirical formula
\begin{equation}
\changedforRone{\tau(\phi,t_{\rm eq})} = A(t_{\rm eq}) + \displaystyle\frac{B(t_{\rm eq})}{[\phi_{\rm J}(t_{\rm eq}) - \phi]^2}
\label{eq:tkphi}
\end{equation}
increase via thermal onset; $\phi_{\rm J}(t_{\rm eq})$ increases roughly logarithmically with $t_{\rm eq}$, from $\sim\!\! 0.648$ to $\sim\!\! 0.674$ over the range $10^1 \leq t_{\rm eq} \leq 10^5$.
\otherchanges{Comparable increases in $\phi_{\rm J}(t_{\rm eq})$ have} been reported before \otherchanges{-- they arise} from relatively-well-understood thermal onset effects \cite{sastry98,debenedetti01,chaudhuri10,ozawa17,morse21} \otherchanges{-- but} the concomitant shift of the ranges of $\phi < \phi_{\rm J}(t_{\rm eq})$ over which relaxation times for energy minimization diverge has not (to the best of our knowledge) been previously reported.

Panel (b) illustrates a closely associated effect. 
When $\phi_{\rm J}(t_{\rm eq}) < \phi$, $ \tau$ values are effectively infinite since systems never unjam.
As thermal onset proceeds, $ \tau$ values become finite (but large) as soon as $\phi_{\rm J}(t_{\rm eq})$ exceeds $\phi$, then drop by $\sim 2$ orders of magnitude as systems approach equilibrium.
Below, we will interpret this result in terms of how thermal soft-sphere glasses' most-typically-occupied PEL basins evolve during constant-$\phi$ aging, and suggest how it might be experimentally characterized.

Panel (c) shows that $ \tau$  diverges with increasing $Z_{\rm ms}( \tau)$ approximately as
\begin{equation}
\changedforRone{\tau(\phi,t_{\rm eq})}  = C + \displaystyle\frac{D}{\left(Z_{\rm iso} - Z_{\rm ms}[\changedforRone{\tau(\phi,t_{\rm eq})}]\right)^2},
\label{eq:tkZnr}
\end{equation}
where $C$ and $D$ are $t_{\rm eq}$-independent constants.
The common inverse-quadratic form of the diverging time scales illustrated in panels (a) and (c) arises rather trivially since  $Z_{\rm ms}( \tau)$ increases linearly with $\phi$ over the range of packing fractions for which Eqs.\ \ref{eq:tkphi}-\ref{eq:tkZnr} describe the data.
On the other hand, the results presented in panel (c) [unlike those of panel (a)] unambiguously show (i) that plotting the terminal relaxation times for thermal glasses' energy minimization as a function of the $Z_{\rm ms}$ values at those times allows one to (nearly) collapse results that look very different when plotted as a function of $\phi$, and (ii) these times always diverge when systems are isostatic \textit{at the top of} the final PEL sub-basin they encounter.

We emphasize that [owing to the concerns raised in Ref.\ \cite{nishikawa21} and the upturns in $\tau$ at small $\Delta Z$ that are visible in Fig.\ \ref{fig:fire2}(c)] we are \textit{not} asserting that the above results imply a true critical phenomenon with $\nu = 2$.
\changedforRone{Employing a different energy-minimization algorithm can change the values of both $\tau$ and $\nu$ \cite{SuppMat,foot2p7,footsmalltauF}}, 
and our goal for this study is not to formulate an exact physics picture for $\tau(\phi, t_{\rm eq})$.
\changedforRone{Instead our goals are to demonstrate that (i) the divergence of $\tau(\phi,t_{\rm eq})$ exhibits thermal onset, and (ii) examining which aspects of the $\tau(\phi,t_{\rm eq})$ data \textit{remain the same} as equilibration/onset proceeds allows one to formulate a simplified picture of these systems' strongly $\phi$- and $t_{\rm eq}$-dependent energy-minimization dynamics.}

\changedforRtwo{For example, the data shown in Fig.\ \ref{fig:fire2}(a) would paint a very confusing picture if the color-coding were removed, or if one attempted to compare results for systems that had been prepared using different equilibration protocols.
In contrast,  Fig.\ 2(c) is much easier to understand.
It shows that the terminal relaxation times for energy minimization in thermalized soft-sphere glasses are always controlled by the the lowest-lying structure of systems' PELs.  
This structure can evolve \textit{dramatically} with equilibration or ``waiting'' time owing to thermal onset, but the collapse illustrated in Fig. 2(c) shows that the effects of this evolution on $\tau(\phi,t_{\rm eq})$ can be understood almost trivially.
This is the central result of our study.
Our demonstration that $Z_{\rm ms}[\tau(\phi,t_{\rm eq})]$ rather than $\phi$ is the relevant control variable for thermal soft-sphere glasses' energy-minimization dynamics might also serve as the basis for a critical-phenomena-based theory for these dynamics, formulated along the lines of Refs.\ \cite{goodrich16,liarte22}.}

All trends \otherchanges{reported above} indicate that systems spend a diverging amount of time near the boundaries between sub-basins that have large $Z$ but very small $E_{\rm p}$, and that they encounter more and more of these boundaries as $\phi \to \phi_{\rm J}(t_{\rm eq})$ from below.
\otherchanges{This is} consistent with the Gardner-like-physics prediction of a proliferation of sub-basins with very small but nonzero energy \otherchanges{\cite{gardner85,charbonneau14}}, and with recent studies suggesting that glasses subjected to thermal quenches spend a diverging amount of time (as $\phi \to \phi_{\rm J}$ from below) traversing saddle points as they explore their PELs and gradually fall into ever-lower sub-basins before finally unjamming \cite{scalliet19,parley22,nishikawa22,manacorda22,olsson22,charbonneau23}.
The proliferation of kinks, since they correspond to changes of direction of $\vec{F}$ and $\vec{v}$, agrees with Ref.\ \cite{hwang16}'s demonstration that systems near jamming follow fractal paths through configuration space during FIRE energy minimization.

\begin{figure}[h]
\includegraphics[width=3.125in]{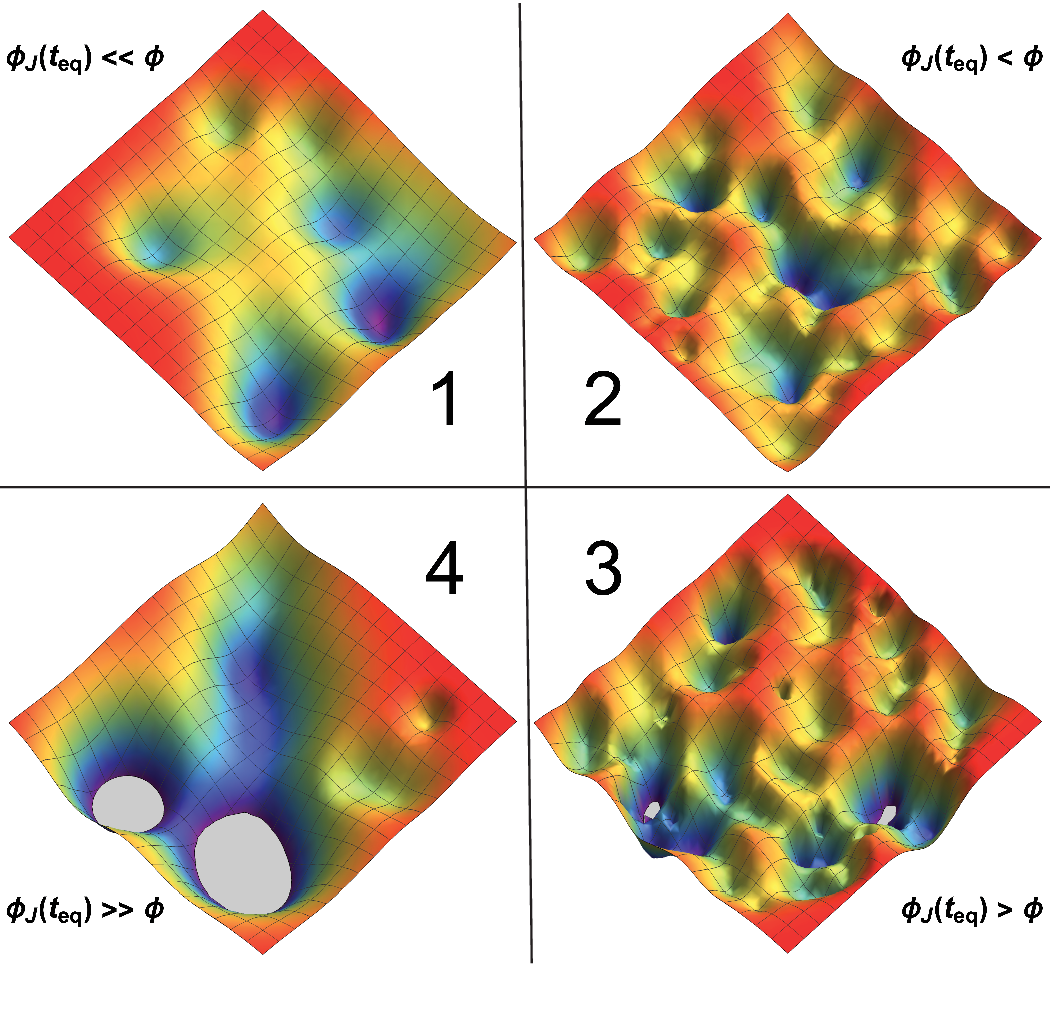}
\caption{\changedforRtwo{Schematic illustration of how the PEL regions most typically sampled by thermal soft-sphere glasses evolve as they age at constant $\phi$ and $T$  following a thermal quench.
Unjammed regions of the PEL are gray-colored.}}
\label{fig:evolutionschem}
\end{figure}

Taken together, our results suggest the following \changedforRtwo{four-stage} picture (schematically illustrated in Figure \ref{fig:evolutionschem}) for how the character of the PEL basins \otherchanges{that simulated dense}  thermal soft-sphere glasses most typically occupy evolves as they are equilibrated at constant $\phi$ and $T$ following a thermal quench:
\changedforRtwo{(1) When $t_{\rm eq}$ is small and $\phi_J(t_{\rm eq})$ remains well below $\phi$, systems are typically in a smooth portion of their PEL and can quickly find the bottom of a nearby jammed basin;
(2) As $\phi_{\rm J}(t_{\rm eq})$ approaches $\phi$ from below, systems cross into rougher portions of their PELs that have more ``wrinkles'' (basin boundaries), and hence take longer to find the bottom of a jammed basin;
(3) As $\phi_{\rm J}(t_{\rm eq})$ increases past $\phi$, unjammed basins emerge, but to reach them, systems must traverse} very rough regions of their PELs characterized by a proliferation of basins with fractal boundaries \cite{charbonneau14,hwang16}; and
\changedforRtwo{(4) Finally, as $t_{\rm eq} \to \infty$ and $\phi_{\rm J}(t_{\rm eq})$ grows further beyond $\phi$, systems cross back into progressively smoother portions of their PELs where they can more quickly find the bottom of a nearby unjammed basin.}

\changedforRtwo{Numerous studies have shown that both athermal and quenched glasses go through this process as $\phi$ decreases from well above to well below $\phi_{\rm RCP}$ \cite{charbonneau14,morse14,morse17,saitoh20,manacorda22,nishikawa22}. 
Here we have argued that it should also occur in sufficiently-dense thermal glasses maintained at \textit{fixed} $\phi$ and $T$, as their thermalized pair energy $E_{\rm p}(t_{\rm eq})$ and pressure  $P(t_{\rm eq})$ slowly decrease via aging \cite{kob97b,mendoza22}.
Since the transition from stage (2) to stage (3) is associated with both diverging length scales \cite{wyart05,lerner12,nishikawa21}
and diverging time scales such as $\tau^*$, $ \tau$, and $\omega_{\rm min}^{-1}$, it should have multiple signatures whose observation does not require energy minimization.
For example, it should also produce a nonmonotonic evolution of systems' low-energy vibrational spectra that should be easily observable in simulations \cite{silbert05,ikeda13} and potentially observable in experiments \cite{footsurov,surovtsev03,sokolov92}.
Finally we point out that stage (4) can be interpreted as an extension of the well-known reduction of $E_{\rm IS}(\phi)$ with increasing $t_{\rm eq}$ \cite{sastry98,debenedetti01}.
While  $E_{\rm IS}(\phi,t_{\rm eq}) = 0$ for all systems that ultimately unjam, i.e.\ for all $t_{\rm eq} > t_{\rm unjam}(\phi)$, thermal onset continues even for $t_{\rm eq} > t_{\rm unjam}(\phi)$, in the sense that systems continue moving into progressively smoother regions of their PELs where the unjammed basins are both larger and more easily accessible.}

We thank Patrick Charbonneau for helpful discussions.
This material is based upon work supported by the National Science Foundation under Grant Nos.\ DMR-2026271 \otherchanges{and DMR-2419261}.


\begin{thebibliography}{61}%
\makeatletter
\providecommand \@ifxundefined [1]{%
 \@ifx{#1\undefined}
}%
\providecommand \@ifnum [1]{%
 \ifnum #1\expandafter \@firstoftwo
 \else \expandafter \@secondoftwo
 \fi
}%
\providecommand \@ifx [1]{%
 \ifx #1\expandafter \@firstoftwo
 \else \expandafter \@secondoftwo
 \fi
}%
\providecommand \natexlab [1]{#1}%
\providecommand \enquote  [1]{``#1''}%
\providecommand \bibnamefont  [1]{#1}%
\providecommand \bibfnamefont [1]{#1}%
\providecommand \citenamefont [1]{#1}%
\providecommand \href@noop [0]{\@secondoftwo}%
\providecommand \href [0]{\begingroup \@sanitize@url \@href}%
\providecommand \@href[1]{\@@startlink{#1}\@@href}%
\providecommand \@@href[1]{\endgroup#1\@@endlink}%
\providecommand \@sanitize@url [0]{\catcode `\\12\catcode `\$12\catcode
  `\&12\catcode `\#12\catcode `\^12\catcode `\_12\catcode `\%12\relax}%
\providecommand \@@startlink[1]{}%
\providecommand \@@endlink[0]{}%
\providecommand \url  [0]{\begingroup\@sanitize@url \@url }%
\providecommand \@url [1]{\endgroup\@href {#1}{\urlprefix }}%
\providecommand \urlprefix  [0]{URL }%
\providecommand \Eprint [0]{\href }%
\providecommand \doibase [0]{http://dx.doi.org/}%
\providecommand \selectlanguage [0]{\@gobble}%
\providecommand \bibinfo  [0]{\@secondoftwo}%
\providecommand \bibfield  [0]{\@secondoftwo}%
\providecommand \translation [1]{[#1]}%
\providecommand \BibitemOpen [0]{}%
\providecommand \bibitemStop [0]{}%
\providecommand \bibitemNoStop [0]{.\EOS\space}%
\providecommand \EOS [0]{\spacefactor3000\relax}%
\providecommand \BibitemShut  [1]{\csname bibitem#1\endcsname}%
\let\auto@bib@innerbib\@empty
\bibitem [{\citenamefont {O'Hern}\ \emph {et~al.}(2003)\citenamefont {O'Hern},
  \citenamefont {Silbert}, \citenamefont {Liu},\ and\ \citenamefont
  {Nagel}}]{ohern03}%
  \BibitemOpen
  \bibfield  {author} {\bibinfo {author} {\bibfnamefont {C.~S.}\ \bibnamefont
  {O'Hern}}, \bibinfo {author} {\bibfnamefont {L.~E.}\ \bibnamefont {Silbert}},
  \bibinfo {author} {\bibfnamefont {A.~J.}\ \bibnamefont {Liu}}, \ and\
  \bibinfo {author} {\bibfnamefont {S.~R.}\ \bibnamefont {Nagel}},\ }\bibfield
  {title} {\enquote {\bibinfo {title} {Jamming at zero temperature and zero
  applied stress: The epitome of disorder},}\ }\href@noop {} {\bibfield
  {journal} {\bibinfo  {journal} {Phys. Rev. E}\ }\textbf {\bibinfo {volume}
  {68}},\ \bibinfo {pages} {011306} (\bibinfo {year} {2003})}\BibitemShut
  {NoStop}%
\bibitem [{\citenamefont {Wyart}\ \emph {et~al.}(2005)\citenamefont {Wyart},
  \citenamefont {Nagel},\ and\ \citenamefont {Witten}}]{wyart05}%
  \BibitemOpen
  \bibfield  {author} {\bibinfo {author} {\bibfnamefont {M.}~\bibnamefont
  {Wyart}}, \bibinfo {author} {\bibfnamefont {S.~R.}\ \bibnamefont {Nagel}}, \
  and\ \bibinfo {author} {\bibfnamefont {T.~A.}\ \bibnamefont {Witten}},\
  }\bibfield  {title} {\enquote {\bibinfo {title} {Geometric origin of excess
  low-frequency vibrational modes in weakly connected amorphous solids},}\
  }\href@noop {} {\bibfield  {journal} {\bibinfo  {journal} {Europhys. Lett.}\
  }\textbf {\bibinfo {volume} {72}},\ \bibinfo {pages} {486} (\bibinfo {year}
  {2005})}\BibitemShut {NoStop}%
\bibitem [{\citenamefont {Keys}\ \emph {et~al.}(2007)\citenamefont {Keys},
  \citenamefont {Abate}, \citenamefont {Glotzer},\ and\ \citenamefont
  {Durian}}]{keys07}%
  \BibitemOpen
  \bibfield  {author} {\bibinfo {author} {\bibfnamefont {A.~S.}\ \bibnamefont
  {Keys}}, \bibinfo {author} {\bibfnamefont {A.~R.}\ \bibnamefont {Abate}},
  \bibinfo {author} {\bibfnamefont {S.~C.}\ \bibnamefont {Glotzer}}, \ and\
  \bibinfo {author} {\bibfnamefont {D.~J.}\ \bibnamefont {Durian}},\ }\bibfield
   {title} {\enquote {\bibinfo {title} {Measurement of growing dynamical length
  scales and prediction of the jamming transition in a granular material},}\
  }\href@noop {} {\bibfield  {journal} {\bibinfo  {journal} {Nature Phys.}\
  }\textbf {\bibinfo {volume} {3}},\ \bibinfo {pages} {260} (\bibinfo {year}
  {2007})}\BibitemShut {NoStop}%
\bibitem [{\citenamefont {Hopkins}\ \emph {et~al.}(2012)\citenamefont
  {Hopkins}, \citenamefont {Stillinger},\ and\ \citenamefont
  {Torquato}}]{hopkins12a}%
  \BibitemOpen
  \bibfield  {author} {\bibinfo {author} {\bibfnamefont {A.~B.}\ \bibnamefont
  {Hopkins}}, \bibinfo {author} {\bibfnamefont {F.~H.}\ \bibnamefont
  {Stillinger}}, \ and\ \bibinfo {author} {\bibfnamefont {S.}~\bibnamefont
  {Torquato}},\ }\bibfield  {title} {\enquote {\bibinfo {title} {Nonequilibrium
  static diverging lengthscales on approaching a prototypical model glassy
  state},}\ }\href@noop {} {\bibfield  {journal} {\bibinfo  {journal} {Phys.
  Rev. E}\ }\textbf {\bibinfo {volume} {86}},\ \bibinfo {pages} {021505}
  (\bibinfo {year} {2012})}\BibitemShut {NoStop}%
\bibitem [{\citenamefont {Berthier}\ \emph {et~al.}(2016)\citenamefont
  {Berthier}, \citenamefont {Charbonneau}, \citenamefont {Jin}, \citenamefont
  {Parisi}, \citenamefont {Seoane},\ and\ \citenamefont
  {Zamponi}}]{berthier16b}%
  \BibitemOpen
  \bibfield  {author} {\bibinfo {author} {\bibfnamefont {L.}~\bibnamefont
  {Berthier}}, \bibinfo {author} {\bibfnamefont {P.}~\bibnamefont
  {Charbonneau}}, \bibinfo {author} {\bibfnamefont {Y.}~\bibnamefont {Jin}},
  \bibinfo {author} {\bibfnamefont {G.}~\bibnamefont {Parisi}}, \bibinfo
  {author} {\bibfnamefont {B.}~\bibnamefont {Seoane}}, \ and\ \bibinfo {author}
  {\bibfnamefont {F.}~\bibnamefont {Zamponi}},\ }\bibfield  {title} {\enquote
  {\bibinfo {title} {Growing timescales and lengthscales characterizing
  vibrations of amorphous solids},}\ }\href@noop {} {\bibfield  {journal}
  {\bibinfo  {journal} {Proc. Nat. Acad. Sci.}\ }\textbf {\bibinfo {volume}
  {113}},\ \bibinfo {pages} {8397} (\bibinfo {year} {2016})}\BibitemShut
  {NoStop}%
\bibitem [{\citenamefont {Brady}(1993)}]{brady93}%
  \BibitemOpen
  \bibfield  {author} {\bibinfo {author} {\bibfnamefont {J.~F.}\ \bibnamefont
  {Brady}},\ }\bibfield  {title} {\enquote {\bibinfo {title} {The rheological
  behavior of concentrated colloidal dispersions},}\ }\href@noop {} {\bibfield
  {journal} {\bibinfo  {journal} {J. Chem. Phys.}\ }\textbf {\bibinfo {volume}
  {99}},\ \bibinfo {pages} {567} (\bibinfo {year} {1993})}\BibitemShut
  {NoStop}%
\bibitem [{\citenamefont {Olsson}\ and\ \citenamefont
  {Teitel}(2007)}]{olsson07}%
  \BibitemOpen
  \bibfield  {author} {\bibinfo {author} {\bibfnamefont {P.}~\bibnamefont
  {Olsson}}\ and\ \bibinfo {author} {\bibfnamefont {S.}~\bibnamefont
  {Teitel}},\ }\bibfield  {title} {\enquote {\bibinfo {title} {Critical scaling
  of shear viscosity at the jamming transition},}\ }\href@noop {} {\bibfield
  {journal} {\bibinfo  {journal} {Phys. Rev. Lett.}\ }\textbf {\bibinfo
  {volume} {99}},\ \bibinfo {pages} {178001} (\bibinfo {year}
  {2007})}\BibitemShut {NoStop}%
\bibitem [{\citenamefont {Forterre}\ and\ \citenamefont
  {Pouliquen}(2008)}]{forterre08}%
  \BibitemOpen
  \bibfield  {author} {\bibinfo {author} {\bibfnamefont {Y.}~\bibnamefont
  {Forterre}}\ and\ \bibinfo {author} {\bibfnamefont {O.}~\bibnamefont
  {Pouliquen}},\ }\bibfield  {title} {\enquote {\bibinfo {title} {Flows of
  dense granular media},}\ }\href@noop {} {\bibfield  {journal} {\bibinfo
  {journal} {Annu. Rev. Fluid Mech.}\ }\textbf {\bibinfo {volume} {40}},\
  \bibinfo {pages} {1} (\bibinfo {year} {2008})}\BibitemShut {NoStop}%
\bibitem [{\citenamefont {Nordstrom}\ \emph {et~al.}(2010)\citenamefont
  {Nordstrom}, \citenamefont {Verneuil}, \citenamefont {Arratia}, \citenamefont
  {Basu}, \citenamefont {Zhang}, \citenamefont {Yodh}, \citenamefont {Gollub},\
  and\ \citenamefont {Durian}}]{nordstrom10}%
  \BibitemOpen
  \bibfield  {author} {\bibinfo {author} {\bibfnamefont {K.~N.}\ \bibnamefont
  {Nordstrom}}, \bibinfo {author} {\bibfnamefont {E.}~\bibnamefont {Verneuil}},
  \bibinfo {author} {\bibfnamefont {P.~E.}\ \bibnamefont {Arratia}}, \bibinfo
  {author} {\bibfnamefont {A.}~\bibnamefont {Basu}}, \bibinfo {author}
  {\bibfnamefont {Z.}~\bibnamefont {Zhang}}, \bibinfo {author} {\bibfnamefont
  {A.~G.}\ \bibnamefont {Yodh}}, \bibinfo {author} {\bibfnamefont {J.~P.}\
  \bibnamefont {Gollub}}, \ and\ \bibinfo {author} {\bibfnamefont {D.~J.}\
  \bibnamefont {Durian}},\ }\bibfield  {title} {\enquote {\bibinfo {title}
  {Microfluidic rheology of soft colloids above and below jamming},}\
  }\href@noop {} {\bibfield  {journal} {\bibinfo  {journal} {Phys. Rev. Lett.}\
  }\textbf {\bibinfo {volume} {105}},\ \bibinfo {pages} {175701} (\bibinfo
  {year} {2010})}\BibitemShut {NoStop}%
\bibitem [{\citenamefont {Boyer}\ \emph {et~al.}(2011)\citenamefont {Boyer},
  \citenamefont {Guazzelli},\ and\ \citenamefont {Pouliquen}}]{boyer11}%
  \BibitemOpen
  \bibfield  {author} {\bibinfo {author} {\bibfnamefont {F.}~\bibnamefont
  {Boyer}}, \bibinfo {author} {\bibfnamefont {E.}~\bibnamefont {Guazzelli}}, \
  and\ \bibinfo {author} {\bibfnamefont {O.}~\bibnamefont {Pouliquen}},\
  }\bibfield  {title} {\enquote {\bibinfo {title} {Unifying suspension and
  granular rheology},}\ }\href@noop {} {\bibfield  {journal} {\bibinfo
  {journal} {Phys. Rev. Lett.}\ }\textbf {\bibinfo {volume} {107}},\ \bibinfo
  {pages} {188301} (\bibinfo {year} {2011})}\BibitemShut {NoStop}%
\bibitem [{\citenamefont {Andreotti}\ \emph {et~al.}(2012)\citenamefont
  {Andreotti}, \citenamefont {Barrat},\ and\ \citenamefont
  {Heussinger}}]{andreotti12}%
  \BibitemOpen
  \bibfield  {author} {\bibinfo {author} {\bibfnamefont {B.}~\bibnamefont
  {Andreotti}}, \bibinfo {author} {\bibfnamefont {{J.-L.}}\ \bibnamefont
  {Barrat}}, \ and\ \bibinfo {author} {\bibfnamefont {C.}~\bibnamefont
  {Heussinger}},\ }\bibfield  {title} {\enquote {\bibinfo {title} {Shear flow
  of non-{B}rownian suspensions close to jamming},}\ }\href@noop {} {\bibfield
  {journal} {\bibinfo  {journal} {Phys. Rev. Lett.}\ }\textbf {\bibinfo
  {volume} {109}},\ \bibinfo {pages} {105901} (\bibinfo {year}
  {2012})}\BibitemShut {NoStop}%
\bibitem [{\citenamefont {Kawasaki}\ \emph {et~al.}(2015)\citenamefont
  {Kawasaki}, \citenamefont {Coslovich}, \citenamefont {Ikeda},\ and\
  \citenamefont {Berthier}}]{kawasaki15}%
  \BibitemOpen
  \bibfield  {author} {\bibinfo {author} {\bibfnamefont {T.}~\bibnamefont
  {Kawasaki}}, \bibinfo {author} {\bibfnamefont {D.}~\bibnamefont {Coslovich}},
  \bibinfo {author} {\bibfnamefont {A.}~\bibnamefont {Ikeda}}, \ and\ \bibinfo
  {author} {\bibfnamefont {L.}~\bibnamefont {Berthier}},\ }\bibfield  {title}
  {\enquote {\bibinfo {title} {Diverging viscosity and soft granular rheology
  in non-Brownian suspensions},}\ }\href@noop {} {\bibfield  {journal}
  {\bibinfo  {journal} {Phys. Rev. E}\ }\textbf {\bibinfo {volume} {91}},\
  \bibinfo {pages} {012203} (\bibinfo {year} {2015})}\BibitemShut {NoStop}%
\bibitem [{\citenamefont {Suzuki}\ and\ \citenamefont
  {Hayakawa}(2015)}]{suzuki15}%
  \BibitemOpen
  \bibfield  {author} {\bibinfo {author} {\bibfnamefont {K.}~\bibnamefont
  {Suzuki}}\ and\ \bibinfo {author} {\bibfnamefont {H.}~\bibnamefont
  {Hayakawa}},\ }\bibfield  {title} {\enquote {\bibinfo {title} {Divergence of
  viscosity in jammed granular materials: A theoretical approach},}\
  }\href@noop {} {\bibfield  {journal} {\bibinfo  {journal} {Phys. Rev. Lett.}\
  }\textbf {\bibinfo {volume} {115}},\ \bibinfo {pages} {098001} (\bibinfo
  {year} {2015})}\BibitemShut {NoStop}%
\bibitem [{\citenamefont {Heussinger}\ and\ \citenamefont
  {Barrat}(2009)}]{heussinger09}%
  \BibitemOpen
  \bibfield  {author} {\bibinfo {author} {\bibfnamefont {C.}~\bibnamefont
  {Heussinger}}\ and\ \bibinfo {author} {\bibfnamefont {J.-L.}\ \bibnamefont
  {Barrat}},\ }\bibfield  {title} {\enquote {\bibinfo {title} {Jamming
  transition as probed by quasistatic shear flow},}\ }\href@noop {} {\bibfield
  {journal} {\bibinfo  {journal} {Phys. Rev. Lett.}\ }\textbf {\bibinfo
  {volume} {102}},\ \bibinfo {pages} {218303} (\bibinfo {year}
  {2009})}\BibitemShut {NoStop}%
\bibitem [{\citenamefont {Olsson}(2019)}]{olsson19}%
  \BibitemOpen
  \bibfield  {author} {\bibinfo {author} {\bibfnamefont {P.}~\bibnamefont
  {Olsson}},\ }\bibfield  {title} {\enquote {\bibinfo {title} {Dimensionality
  and viscosity exponent in shear-driven jamming},}\ }\href@noop {} {\bibfield
  {journal} {\bibinfo  {journal} {Phys. Rev. Lett.}\ }\textbf {\bibinfo
  {volume} {122}},\ \bibinfo {pages} {108003} (\bibinfo {year}
  {2019})}\BibitemShut {NoStop}%
\bibitem [{\citenamefont {Lerner}\ \emph
  {et~al.}(2012{\natexlab{a}})\citenamefont {Lerner}, \citenamefont
  {D{\"u}ring},\ and\ \citenamefont {Wyart}}]{lerner12}%
  \BibitemOpen
  \bibfield  {author} {\bibinfo {author} {\bibfnamefont {E.}~\bibnamefont
  {Lerner}}, \bibinfo {author} {\bibfnamefont {G.}~\bibnamefont {D{\"u}ring}},
  \ and\ \bibinfo {author} {\bibfnamefont {M.}~\bibnamefont {Wyart}},\
  }\bibfield  {title} {\enquote {\bibinfo {title} {A unified framework for
  non-Brownian suspension flows and soft amorphous solids},}\ }\href@noop {}
  {\bibfield  {journal} {\bibinfo  {journal} {Proc. Nat. Acad. Sci.}\ }\textbf
  {\bibinfo {volume} {109}},\ \bibinfo {pages} {4798} (\bibinfo {year}
  {2012}{\natexlab{a}})}\BibitemShut {NoStop}%
\bibitem [{\citenamefont {Lerner}\ \emph
  {et~al.}(2012{\natexlab{b}})\citenamefont {Lerner}, \citenamefont
  {D{\"u}ring},\ and\ \citenamefont {Wyart}}]{lerner12b}%
  \BibitemOpen
  \bibfield  {author} {\bibinfo {author} {\bibfnamefont {E.}~\bibnamefont
  {Lerner}}, \bibinfo {author} {\bibfnamefont {G.}~\bibnamefont {D{\"u}ring}},
  \ and\ \bibinfo {author} {\bibfnamefont {M.}~\bibnamefont {Wyart}},\
  }\bibfield  {title} {\enquote {\bibinfo {title} {Toward a microscopic
  description of flow near the jamming threshold},}\ }\href@noop {} {\bibfield
  {journal} {\bibinfo  {journal} {Europhysics Letters}\ }\textbf {\bibinfo
  {volume} {99}},\ \bibinfo {pages} {58003} (\bibinfo {year}
  {2012}{\natexlab{b}})}\BibitemShut {NoStop}%
\bibitem [{\citenamefont {Olsson}(2015)}]{olsson15}%
  \BibitemOpen
  \bibfield  {author} {\bibinfo {author} {\bibfnamefont {P.}~\bibnamefont
  {Olsson}},\ }\bibfield  {title} {\enquote {\bibinfo {title} {Relaxation times
  and rheology in dense athermal suspensions},}\ }\href@noop {} {\bibfield
  {journal} {\bibinfo  {journal} {Phys. Rev. E}\ }\textbf {\bibinfo {volume}
  {91}},\ \bibinfo {pages} {062209} (\bibinfo {year} {2015})}\BibitemShut
  {NoStop}%
\bibitem [{\citenamefont {Ikeda}\ \emph {et~al.}(2020)\citenamefont {Ikeda},
  \citenamefont {Kawasaki}, \citenamefont {Berthier}, \citenamefont {Saitoh},\
  and\ \citenamefont {Hatano}}]{ikeda20}%
  \BibitemOpen
  \bibfield  {author} {\bibinfo {author} {\bibfnamefont {A.}~\bibnamefont
  {Ikeda}}, \bibinfo {author} {\bibfnamefont {T.}~\bibnamefont {Kawasaki}},
  \bibinfo {author} {\bibfnamefont {L.}~\bibnamefont {Berthier}}, \bibinfo
  {author} {\bibfnamefont {K.}~\bibnamefont {Saitoh}}, \ and\ \bibinfo {author}
  {\bibfnamefont {T.}~\bibnamefont {Hatano}},\ }\bibfield  {title} {\enquote
  {\bibinfo {title} {Universal relaxation dynamics of sphere packings below
  jamming},}\ }\href@noop {} {\bibfield  {journal} {\bibinfo  {journal} {Phys.
  Rev. Lett.}\ }\textbf {\bibinfo {volume} {124}},\ \bibinfo {pages} {058001}
  (\bibinfo {year} {2020})}\BibitemShut {NoStop}%
\bibitem [{\citenamefont {Saitoh}\ \emph {et~al.}(2020)\citenamefont {Saitoh},
  \citenamefont {Hatano}, \citenamefont {Ikeda},\ and\ \citenamefont
  {Tighe}}]{saitoh20}%
  \BibitemOpen
  \bibfield  {author} {\bibinfo {author} {\bibfnamefont {K.}~\bibnamefont
  {Saitoh}}, \bibinfo {author} {\bibfnamefont {T.}~\bibnamefont {Hatano}},
  \bibinfo {author} {\bibfnamefont {A.}~\bibnamefont {Ikeda}}, \ and\ \bibinfo
  {author} {\bibfnamefont {B.~P.}\ \bibnamefont {Tighe}},\ }\bibfield  {title}
  {\enquote {\bibinfo {title} {Stress relaxation above and below the jamming
  transition},}\ }\href@noop {} {\bibfield  {journal} {\bibinfo  {journal}
  {Phys. Rev. Lett.}\ }\textbf {\bibinfo {volume} {124}},\ \bibinfo {pages}
  {118001} (\bibinfo {year} {2020})}\BibitemShut {NoStop}%
\bibitem [{\citenamefont {Nishikawa}\ \emph {et~al.}(2021)\citenamefont
  {Nishikawa}, \citenamefont {Ikeda},\ and\ \citenamefont
  {Berthier}}]{nishikawa21}%
  \BibitemOpen
  \bibfield  {author} {\bibinfo {author} {\bibfnamefont {Y.}~\bibnamefont
  {Nishikawa}}, \bibinfo {author} {\bibfnamefont {A.}~\bibnamefont {Ikeda}}, \
  and\ \bibinfo {author} {\bibfnamefont {L.}~\bibnamefont {Berthier}},\
  }\bibfield  {title} {\enquote {\bibinfo {title} {Relaxation {{Dynamics}} of
  {{Non-Brownian Spheres Below Jamming}}},}\ }\href@noop {} {\bibfield
  {journal} {\bibinfo  {journal} {J. Stat. Phys.}\ }\textbf {\bibinfo {volume}
  {182}},\ \bibinfo {pages} {37} (\bibinfo {year} {2021})}\BibitemShut
  {NoStop}%
\bibitem [{foo({\natexlab{a}})}]{footcalcDOS2}%
  \BibitemOpen
  \bibinfo {note} {\changedforRone{Below jamming, one has to use momentum
  transfer to model the density of states $D(\omega)$. This is usually done
  using either an effective potential or a Fourier transform of the velocity
  auto-correlation function.}}\BibitemShut {Stop}%
\bibitem [{\citenamefont {Torquato}\ \emph {et~al.}(2000)\citenamefont
  {Torquato}, \citenamefont {Truskett},\ and\ \citenamefont
  {Debenedetti}}]{torquato00}%
  \BibitemOpen
  \bibfield  {author} {\bibinfo {author} {\bibfnamefont {S.}~\bibnamefont
  {Torquato}}, \bibinfo {author} {\bibfnamefont {T.~M.}\ \bibnamefont
  {Truskett}}, \ and\ \bibinfo {author} {\bibfnamefont {P.~G.}\ \bibnamefont
  {Debenedetti}},\ }\bibfield  {title} {\enquote {\bibinfo {title} {Is random
  close packing of spheres well defined?}}\ }\href@noop {} {\bibfield
  {journal} {\bibinfo  {journal} {Phys. Rev. Lett.}\ }\textbf {\bibinfo
  {volume} {84}},\ \bibinfo {pages} {2064} (\bibinfo {year}
  {2000})}\BibitemShut {NoStop}%
\bibitem [{\citenamefont {Chaudhuri}\ \emph {et~al.}(2010)\citenamefont
  {Chaudhuri}, \citenamefont {Berthier},\ and\ \citenamefont
  {Sastry}}]{chaudhuri10}%
  \BibitemOpen
  \bibfield  {author} {\bibinfo {author} {\bibfnamefont {P.}~\bibnamefont
  {Chaudhuri}}, \bibinfo {author} {\bibfnamefont {L.}~\bibnamefont {Berthier}},
  \ and\ \bibinfo {author} {\bibfnamefont {S.}~\bibnamefont {Sastry}},\
  }\bibfield  {title} {\enquote {\bibinfo {title} {Jamming transitions in
  amorphous packings of frictionless spheres occur over a continuous range of
  volume fractions},}\ }\href@noop {} {\bibfield  {journal} {\bibinfo
  {journal} {Phys. Rev. Lett.}\ }\textbf {\bibinfo {volume} {104}},\ \bibinfo
  {pages} {165701} (\bibinfo {year} {2010})}\BibitemShut {NoStop}%
\bibitem [{\citenamefont {Ozawa}\ \emph {et~al.}(2017)\citenamefont {Ozawa},
  \citenamefont {Berthier},\ and\ \citenamefont {Coslovich}}]{ozawa17}%
  \BibitemOpen
  \bibfield  {author} {\bibinfo {author} {\bibfnamefont {M.}~\bibnamefont
  {Ozawa}}, \bibinfo {author} {\bibfnamefont {L.}~\bibnamefont {Berthier}}, \
  and\ \bibinfo {author} {\bibfnamefont {D.}~\bibnamefont {Coslovich}},\
  }\bibfield  {title} {\enquote {\bibinfo {title} {Exploring the jamming
  transition over a wide range of critical densities},}\ }\href@noop {}
  {\bibfield  {journal} {\bibinfo  {journal} {SciPost Phys.}\ }\textbf
  {\bibinfo {volume} {3}},\ \bibinfo {pages} {027} (\bibinfo {year}
  {2017})}\BibitemShut {NoStop}%
\bibitem [{\citenamefont {Charbonneau}\ and\ \citenamefont
  {Morse}(2021)}]{morse21}%
  \BibitemOpen
  \bibfield  {author} {\bibinfo {author} {\bibfnamefont {P.}~\bibnamefont
  {Charbonneau}}\ and\ \bibinfo {author} {\bibfnamefont {P.~K.}\ \bibnamefont
  {Morse}},\ }\bibfield  {title} {\enquote {\bibinfo {title} {Memory formation
  in jammed hard spheres},}\ }\href@noop {} {\bibfield  {journal} {\bibinfo
  {journal} {Phys. Rev. Lett.}\ }\textbf {\bibinfo {volume} {126}},\ \bibinfo
  {pages} {088001} (\bibinfo {year} {2021})}\BibitemShut {NoStop}%
\bibitem [{\citenamefont {Sastry}\ \emph {et~al.}(1998)\citenamefont {Sastry},
  \citenamefont {Debenedetti}, \citenamefont {Torquato},\ and\ \citenamefont
  {Stillinger}}]{sastry98}%
  \BibitemOpen
  \bibfield  {author} {\bibinfo {author} {\bibfnamefont {S.}~\bibnamefont
  {Sastry}}, \bibinfo {author} {\bibfnamefont {P.~G.}\ \bibnamefont
  {Debenedetti}}, \bibinfo {author} {\bibfnamefont {S.}~\bibnamefont
  {Torquato}}, \ and\ \bibinfo {author} {\bibfnamefont {F.~H.}\ \bibnamefont
  {Stillinger}},\ }\bibfield  {title} {\enquote {\bibinfo {title} {Signatures
  of distinct dynamical regimes in the energy landscape of a glass-forming
  liquid},}\ }\href@noop {} {\bibfield  {journal} {\bibinfo  {journal}
  {Nature}\ }\textbf {\bibinfo {volume} {393}},\ \bibinfo {pages} {554}
  (\bibinfo {year} {1998})}\BibitemShut {NoStop}%
\bibitem [{\citenamefont {Debenedetti}\ and\ \citenamefont
  {Stillinger}(2001)}]{debenedetti01}%
  \BibitemOpen
  \bibfield  {author} {\bibinfo {author} {\bibfnamefont {P.~G.}\ \bibnamefont
  {Debenedetti}}\ and\ \bibinfo {author} {\bibfnamefont {F.~H.}\ \bibnamefont
  {Stillinger}},\ }\bibfield  {title} {\enquote {\bibinfo {title} {Supercooled
  liquids and the glass transition},}\ }\href@noop {} {\bibfield  {journal}
  {\bibinfo  {journal} {Nature}\ }\textbf {\bibinfo {volume} {410}},\ \bibinfo
  {pages} {259} (\bibinfo {year} {2001})}\BibitemShut {NoStop}%
\bibitem [{\citenamefont {Grigera}\ and\ \citenamefont
  {Parisi}(2001)}]{grigera01}%
  \BibitemOpen
  \bibfield  {author} {\bibinfo {author} {\bibfnamefont {T.~S.}\ \bibnamefont
  {Grigera}}\ and\ \bibinfo {author} {\bibfnamefont {G.}~\bibnamefont
  {Parisi}},\ }\bibfield  {title} {\enquote {\bibinfo {title} {Fast Monte Carlo
  algorithm for supercooled soft spheres},}\ }\href@noop {} {\bibfield
  {journal} {\bibinfo  {journal} {Phys. Rev. E}\ }\textbf {\bibinfo {volume}
  {63}},\ \bibinfo {pages} {045102(R)} (\bibinfo {year} {2001})}\BibitemShut
  {NoStop}%
\bibitem [{\citenamefont {Ninarello}\ \emph {et~al.}(2017)\citenamefont
  {Ninarello}, \citenamefont {Berthier},\ and\ \citenamefont
  {Coslovich}}]{ninarello17}%
  \BibitemOpen
  \bibfield  {author} {\bibinfo {author} {\bibfnamefont {A.}~\bibnamefont
  {Ninarello}}, \bibinfo {author} {\bibfnamefont {L.}~\bibnamefont {Berthier}},
  \ and\ \bibinfo {author} {\bibfnamefont {D.}~\bibnamefont {Coslovich}},\
  }\bibfield  {title} {\enquote {\bibinfo {title} {Models and algorithms for
  the next generation of glass transition studies},}\ }\href@noop {} {\bibfield
   {journal} {\bibinfo  {journal} {Phys. Rev. X}\ }\textbf {\bibinfo {volume}
  {7}},\ \bibinfo {pages} {021039} (\bibinfo {year} {2017})}\BibitemShut
  {NoStop}%
\bibitem [{\citenamefont {Bitzek}\ \emph {et~al.}(2006)\citenamefont {Bitzek},
  \citenamefont {Koskinen}, \citenamefont {G{\"a}hler}, \citenamefont
  {Moseler},\ and\ \citenamefont {Gumbsch}}]{bitzek06}%
  \BibitemOpen
  \bibfield  {author} {\bibinfo {author} {\bibfnamefont {E.}~\bibnamefont
  {Bitzek}}, \bibinfo {author} {\bibfnamefont {P.}~\bibnamefont {Koskinen}},
  \bibinfo {author} {\bibfnamefont {F.}~\bibnamefont {G{\"a}hler}}, \bibinfo
  {author} {\bibfnamefont {M.}~\bibnamefont {Moseler}}, \ and\ \bibinfo
  {author} {\bibfnamefont {P.}~\bibnamefont {Gumbsch}},\ }\bibfield  {title}
  {\enquote {\bibinfo {title} {Structural relaxation made simple},}\
  }\href@noop {} {\bibfield  {journal} {\bibinfo  {journal} {Phys. Rev. Lett.}\
  }\textbf {\bibinfo {volume} {97}},\ \bibinfo {pages} {170201} (\bibinfo
  {year} {2006})}\BibitemShut {NoStop}%
\bibitem [{\citenamefont {Gu{\'e}nol{\'e}}\ \emph {et~al.}(2020)\citenamefont
  {Gu{\'e}nol{\'e}}, \citenamefont {N{\"o}hring}, \citenamefont {Vaid},
  \citenamefont {Houll{\'e}}, \citenamefont {Xie}, \citenamefont {Prakash},\
  and\ \citenamefont {Bitzek}}]{guenole20}%
  \BibitemOpen
  \bibfield  {author} {\bibinfo {author} {\bibfnamefont {J.}~\bibnamefont
  {Gu{\'e}nol{\'e}}}, \bibinfo {author} {\bibfnamefont {W.~G.}\ \bibnamefont
  {N{\"o}hring}}, \bibinfo {author} {\bibfnamefont {A.}~\bibnamefont {Vaid}},
  \bibinfo {author} {\bibfnamefont {F.}~\bibnamefont {Houll{\'e}}}, \bibinfo
  {author} {\bibfnamefont {Z.}~\bibnamefont {Xie}}, \bibinfo {author}
  {\bibfnamefont {A.}~\bibnamefont {Prakash}}, \ and\ \bibinfo {author}
  {\bibfnamefont {E.}~\bibnamefont {Bitzek}},\ }\bibfield  {title} {\enquote
  {\bibinfo {title} {Assessment and optimization of the fast inertial
  relaxation engine (FIRE) for energy minimization in atomistic simulations and
  its implementation in LAMMPS},}\ }\href@noop {} {\bibfield  {journal}
  {\bibinfo  {journal} {Comp. Mat. Sci.}\ }\textbf {\bibinfo {volume} {175}},\
  \bibinfo {pages} {109584} (\bibinfo {year} {2020})}\BibitemShut {NoStop}%
\bibitem [{\citenamefont {Hoy}\ and\ \citenamefont
  {Interiano-Alberto}(2022)}]{hoy22}%
  \BibitemOpen
  \bibfield  {author} {\bibinfo {author} {\bibfnamefont {R.~S.}\ \bibnamefont
  {Hoy}}\ and\ \bibinfo {author} {\bibfnamefont {K.~A.}\ \bibnamefont
  {Interiano-Alberto}},\ }\bibfield  {title} {\enquote {\bibinfo {title}
  {Efficient d-dimensional molecular dynamics simulations for studies of the
  glass-jamming transition},}\ }\href@noop {} {\bibfield  {journal} {\bibinfo
  {journal} {Phys. Rev. E}\ }\textbf {\bibinfo {volume} {105}},\ \bibinfo
  {pages} {055305} (\bibinfo {year} {2022})}\BibitemShut {NoStop}%
\bibitem [{Sup()}]{SuppMat}%
  \BibitemOpen
  \bibinfo {note} {\otherchanges{See the Supplemental Material at
  http://link.aps.org/supplemental/XXX/YYY, which includes Refs.\ \cite{shen12,schmiedeberg11},
  for further details on our:\ (1) interparticle interactions; (2)
  implementation of the SWAP and FIRE algorithms; (3) results for the $t_{\rm
  eq}$-dependence of $E_{\rm p}(t)$, $Z_{\rm ms}(t)$, $\phi_{\rm J}$, $A$, and
  $B$; (4) demonstration of the relation of $\tau$ to $\tau^*$; and (5) a
  comparison to results obtained from steepest-descent
  minimization.}}\BibitemShut {Stop}%
\bibitem [{\citenamefont {Shen}\ \emph {et~al.}(2012)\citenamefont {Shen},
  \citenamefont {Schreck}, \citenamefont {Chakraborty}, \citenamefont {Freed},\
  and\ \citenamefont {O'Hern}}]{shen12}%
  \BibitemOpen
  \bibfield  {author} {\bibinfo {author} {\bibfnamefont {T.}~\bibnamefont
  {Shen}}, \bibinfo {author} {\bibfnamefont {C.}~\bibnamefont {Schreck}},
  \bibinfo {author} {\bibfnamefont {B.}~\bibnamefont {Chakraborty}}, \bibinfo
  {author} {\bibfnamefont {D.~E.}\ \bibnamefont {Freed}}, \ and\ \bibinfo
  {author} {\bibfnamefont {C.~S.}\ \bibnamefont {O'Hern}},\ }\bibfield  {title}
  {\enquote {\bibinfo {title} {Structural relaxation in dense liquids composed
  of anisotropic particles},}\ }\href@noop {} {\bibfield  {journal} {\bibinfo
  {journal} {Phys. Rev. E}\ }\textbf {\bibinfo {volume} {86}},\ \bibinfo
  {pages} {041303} (\bibinfo {year} {2012})}\BibitemShut {NoStop}%
\bibitem [{\citenamefont {Schmiedeberg}\ \emph {et~al.}(2011)\citenamefont
  {Schmiedeberg}, \citenamefont {Haxton}, \citenamefont {Nagel},\ and\
  \citenamefont {Liu}}]{schmiedeberg11}%
  \BibitemOpen
  \bibfield  {author} {\bibinfo {author} {\bibfnamefont {M.}~\bibnamefont
  {Schmiedeberg}}, \bibinfo {author} {\bibfnamefont {T.~K.}\ \bibnamefont
  {Haxton}}, \bibinfo {author} {\bibfnamefont {S.~R.}\ \bibnamefont {Nagel}}, \
  and\ \bibinfo {author} {\bibfnamefont {A.~J.}\ \bibnamefont {Liu}},\
  }\bibfield  {title} {\enquote {\bibinfo {title} {Mapping the glassy dynamics
  of soft spheres onto hard-sphere behavior},}\ }\href@noop {} {\bibfield
  {journal} {\bibinfo  {journal} {Europhys. Lett.}\ }\textbf {\bibinfo {volume}
  {96}},\ \bibinfo {pages} {36010} (\bibinfo {year} {2011})}\BibitemShut
  {NoStop}%
\bibitem [{foo({\natexlab{b}})}]{footd23}%
  \BibitemOpen
  \bibinfo {note} {\changedforRtwo{To suppress any fractionation, we employ a
  polydispersity index ($\Delta = 0.107$) that is lower than that employed in
  several recent studies. For example, Refs. \cite{ninarello17,ozawa17} used
  the same $P(\sigma) \sim \sigma^{-3}$ \cite{SuppMat}, but with $\Delta =
  0.23$.}}\BibitemShut {Stop}%
\bibitem [{foo({\natexlab{c}})}]{footZdp1}%
  \BibitemOpen
  \bibinfo {note} {The $\changedforRtwo{Z_i} \geq d+1$ criterion was used to
  identify \otherchanges{non-}rattlers in Refs.\
  \cite{lerner12,lerner12b,olsson15,olsson19,ikeda20,saitoh20,nishikawa21}, but
  in contrast to these studies, we do \textit{not} iteratively remove particles
  with $d+1$ contacts prior to calculating the final $Z$ values.}\BibitemShut
  {Stop}%
\bibitem [{\citenamefont {Morse}\ and\ \citenamefont {Corwin}(2023)}]{morse23}%
  \BibitemOpen
  \bibfield  {author} {\bibinfo {author} {\bibfnamefont {P.~K.}\ \bibnamefont
  {Morse}}\ and\ \bibinfo {author} {\bibfnamefont {E.~I.}\ \bibnamefont
  {Corwin}},\ }\bibfield  {title} {\enquote {\bibinfo {title} {Local stability
  of spheres via the convex hull and the radical Voronoi diagram},}\
  }\href@noop {} {\bibfield  {journal} {\bibinfo  {journal} {Phys. Rev. E}\
  }\textbf {\bibinfo {volume} {108}},\ \bibinfo {pages} {064901} (\bibinfo
  {year} {2023})}\BibitemShut {NoStop}%
\bibitem [{foo({\natexlab{d}})}]{foot2p7}%
  \BibitemOpen
  \bibinfo {note} {The larger $\nu$ reported in Refs.\
  \cite{lerner12,lerner12b,olsson15,ikeda20,saitoh20,olsson19,nishikawa21} may
  arise from the different minimization algorithm they employed (i.e.\ SD
  energy minimization produces overdamped dynamics whereas FIRE produces damped
  inertial dynamics), and/or from their different definition of $Z_{\rm ms}$
  \cite{nishikawa21}. In general, $\tau^* = \omega_{\rm min}^{-2}/2$ is
  expected only for \textit{overdamped} minimization dynamics; inertial
  dynamics give $\tau^* \sim \omega_{\rm min}^{-1}$.}\BibitemShut {Stop}%
\bibitem [{foo({\natexlab{e}})}]{footsmalltauF}%
  \BibitemOpen
  \bibinfo {note} {Our $\tau$ values are \otherchanges{one} or more orders of
  magnitude smaller that the corresponding times for SD minimization for the
  same systems, and indeed smaller than the $\tau^*$ values reported in Refs.\
  \cite{lerner12,lerner12b,olsson15,olsson19,ikeda20,saitoh20,nishikawa21} for
  systems at comparable $\Delta \phi$, owing to FIRE's more efficient
  implementation.}\BibitemShut {Stop}%
\bibitem [{\citenamefont {Olsson}(2022)}]{olsson22}%
  \BibitemOpen
  \bibfield  {author} {\bibinfo {author} {\bibfnamefont {P.}~\bibnamefont
  {Olsson}},\ }\bibfield  {title} {\enquote {\bibinfo {title} {Relaxation
  times, rheology, and finite size effects for non-{Brownian} disks in two
  dimensions},}\ }\href@noop {} {\bibfield  {journal} {\bibinfo  {journal}
  {Phys. Rev. E}\ }\textbf {\bibinfo {volume} {105}},\ \bibinfo {pages}
  {034902} (\bibinfo {year} {2022})}\BibitemShut {NoStop}%
\bibitem [{\citenamefont {Nishikawa}\ \emph {et~al.}(2022)\citenamefont
  {Nishikawa}, \citenamefont {Ozawa}, \citenamefont {Ikeda}, \citenamefont
  {Chaudhuri},\ and\ \citenamefont {Berthier}}]{nishikawa22}%
  \BibitemOpen
  \bibfield  {author} {\bibinfo {author} {\bibfnamefont {Y.}~\bibnamefont
  {Nishikawa}}, \bibinfo {author} {\bibfnamefont {M.}~\bibnamefont {Ozawa}},
  \bibinfo {author} {\bibfnamefont {A.}~\bibnamefont {Ikeda}}, \bibinfo
  {author} {\bibfnamefont {P.}~\bibnamefont {Chaudhuri}}, \ and\ \bibinfo
  {author} {\bibfnamefont {L.}~\bibnamefont {Berthier}},\ }\bibfield  {title}
  {\enquote {\bibinfo {title} {Relazation dynamics in the energy landscape of
  glass-forming liquids},}\ }\href@noop {} {\bibfield  {journal} {\bibinfo
  {journal} {Phys. Rev. X}\ }\textbf {\bibinfo {volume} {12}},\ \bibinfo
  {pages} {021001} (\bibinfo {year} {2022})}\BibitemShut {NoStop}%
\bibitem [{\citenamefont {Manacorda}\ and\ \citenamefont
  {Zamponi}(2022)}]{manacorda22}%
  \BibitemOpen
  \bibfield  {author} {\bibinfo {author} {\bibfnamefont {A.}~\bibnamefont
  {Manacorda}}\ and\ \bibinfo {author} {\bibfnamefont {F.}~\bibnamefont
  {Zamponi}},\ }\bibfield  {title} {\enquote {\bibinfo {title} {Gradient
  descent dynamics and the jamming transition in infinite dimensions},}\
  }\href@noop {} {\bibfield  {journal} {\bibinfo  {journal} {J. Phys. A: Math.
  Theor.}\ }\textbf {\bibinfo {volume} {55}},\ \bibinfo {pages} {224001}
  (\bibinfo {year} {2022})}\BibitemShut {NoStop}%
\bibitem [{\citenamefont {Charbonneau}\ and\ \citenamefont
  {Morse}(2023)}]{charbonneau23}%
  \BibitemOpen
  \bibfield  {author} {\bibinfo {author} {\bibfnamefont {P.}~\bibnamefont
  {Charbonneau}}\ and\ \bibinfo {author} {\bibfnamefont {P.~K.}\ \bibnamefont
  {Morse}},\ }\bibfield  {title} {\enquote {\bibinfo {title} {Jamming,
  relaxation, and memory in a minimally structured glass former},}\ }\href@noop
  {} {\bibfield  {journal} {\bibinfo  {journal} {Phys. Rev. E}\ }\textbf
  {\bibinfo {volume} {108}},\ \bibinfo {pages} {054102} (\bibinfo {year}
  {2023})}\BibitemShut {NoStop}%
\bibitem [{\citenamefont {Goodrich}\ \emph {et~al.}(2016)\citenamefont
  {Goodrich}, \citenamefont {Liu},\ and\ \citenamefont {Sethna}}]{goodrich16}%
  \BibitemOpen
  \bibfield  {author} {\bibinfo {author} {\bibfnamefont {C.~P.}\ \bibnamefont
  {Goodrich}}, \bibinfo {author} {\bibfnamefont {A.~J.}\ \bibnamefont {Liu}}, \
  and\ \bibinfo {author} {\bibfnamefont {J.~P.}\ \bibnamefont {Sethna}},\
  }\bibfield  {title} {\enquote {\bibinfo {title} {Scaling ansatz for the
  jamming transition},}\ }\href@noop {} {\bibfield  {journal} {\bibinfo
  {journal} {Proc. Nat. Acad. Sci.}\ }\textbf {\bibinfo {volume} {113}},\
  \bibinfo {pages} {9745} (\bibinfo {year} {2016})}\BibitemShut {NoStop}%
\bibitem [{\citenamefont {Liarte}\ \emph {et~al.}(2022)\citenamefont {Liarte},
  \citenamefont {Thornton}, \citenamefont {Schwen}, \citenamefont {Cohen},
  \citenamefont {Chowdhury},\ and\ \citenamefont {Sethna}}]{liarte22}%
  \BibitemOpen
  \bibfield  {author} {\bibinfo {author} {\bibfnamefont {D.~B.}\ \bibnamefont
  {Liarte}}, \bibinfo {author} {\bibfnamefont {S.~J.}\ \bibnamefont
  {Thornton}}, \bibinfo {author} {\bibfnamefont {E.}~\bibnamefont {Schwen}},
  \bibinfo {author} {\bibfnamefont {I.}~\bibnamefont {Cohen}}, \bibinfo
  {author} {\bibfnamefont {D.}~\bibnamefont {Chowdhury}}, \ and\ \bibinfo
  {author} {\bibfnamefont {J.~P.}\ \bibnamefont {Sethna}},\ }\bibfield  {title}
  {\enquote {\bibinfo {title} {Universal scaling for disordered viscoelastic
  matter near the onset of rigidity},}\ }\href@noop {} {\bibfield  {journal}
  {\bibinfo  {journal} {Phys. Rev. E}\ }\textbf {\bibinfo {volume} {106}},\
  \bibinfo {pages} {L052601} (\bibinfo {year} {2022})}\BibitemShut {NoStop}%
\bibitem [{\citenamefont {Gardner}(1985)}]{gardner85}%
  \BibitemOpen
  \bibfield  {author} {\bibinfo {author} {\bibfnamefont {E.}~\bibnamefont
  {Gardner}},\ }\bibfield  {title} {\enquote {\bibinfo {title} {Spin glasses
  with p-spin interactions},}\ }\href@noop {} {\bibfield  {journal} {\bibinfo
  {journal} {Nuc. Phys. B}\ }\textbf {\bibinfo {volume} {257}},\ \bibinfo
  {pages} {747} (\bibinfo {year} {1985})}\BibitemShut {NoStop}%
\bibitem [{\citenamefont {Charbonneau}\ \emph {et~al.}(2014)\citenamefont
  {Charbonneau}, \citenamefont {Kurchan}, \citenamefont {Parisi}, \citenamefont
  {Urbani},\ and\ \citenamefont {Zamponi}}]{charbonneau14}%
  \BibitemOpen
  \bibfield  {author} {\bibinfo {author} {\bibfnamefont {P.}~\bibnamefont
  {Charbonneau}}, \bibinfo {author} {\bibfnamefont {J.}~\bibnamefont
  {Kurchan}}, \bibinfo {author} {\bibfnamefont {G.}~\bibnamefont {Parisi}},
  \bibinfo {author} {\bibfnamefont {P.}~\bibnamefont {Urbani}}, \ and\ \bibinfo
  {author} {\bibfnamefont {F.}~\bibnamefont {Zamponi}},\ }\bibfield  {title}
  {\enquote {\bibinfo {title} {Fractal free energy landscapes in structural
  glasses},}\ }\href@noop {} {\bibfield  {journal} {\bibinfo  {journal} {Nat.
  Comm.}\ }\textbf {\bibinfo {volume} {5}},\ \bibinfo {pages} {3725} (\bibinfo
  {year} {2014})}\BibitemShut {NoStop}%
\bibitem [{\citenamefont {Scalliet}\ \emph {et~al.}(2019)\citenamefont
  {Scalliet}, \citenamefont {Berthier},\ and\ \citenamefont
  {Zamponi}}]{scalliet19}%
  \BibitemOpen
  \bibfield  {author} {\bibinfo {author} {\bibfnamefont {C.}~\bibnamefont
  {Scalliet}}, \bibinfo {author} {\bibfnamefont {L.}~\bibnamefont {Berthier}},
  \ and\ \bibinfo {author} {\bibfnamefont {F.}~\bibnamefont {Zamponi}},\
  }\bibfield  {title} {\enquote {\bibinfo {title} {Nature of excitations and
  defects in structural glasses},}\ }\href@noop {} {\bibfield  {journal}
  {\bibinfo  {journal} {Nature Comm.}\ }\textbf {\bibinfo {volume} {10}},\
  \bibinfo {pages} {5102} (\bibinfo {year} {2019})}\BibitemShut {NoStop}%
\bibitem [{\citenamefont {Parley}\ \emph {et~al.}(2022)\citenamefont {Parley},
  \citenamefont {Mandal},\ and\ \citenamefont {Sollich}}]{parley22}%
  \BibitemOpen
  \bibfield  {author} {\bibinfo {author} {\bibfnamefont {J.~T.}\ \bibnamefont
  {Parley}}, \bibinfo {author} {\bibfnamefont {R.}~\bibnamefont {Mandal}}, \
  and\ \bibinfo {author} {\bibfnamefont {P.}~\bibnamefont {Sollich}},\
  }\bibfield  {title} {\enquote {\bibinfo {title} {Mean-field description of
  aging linear response in athermal amorphous solids},}\ }\href@noop {}
  {\bibfield  {journal} {\bibinfo  {journal} {Phys. Rev. Mater.}\ }\textbf
  {\bibinfo {volume} {6}},\ \bibinfo {pages} {065601} (\bibinfo {year}
  {2022})}\BibitemShut {NoStop}%
\bibitem [{\citenamefont {Hwang}\ \emph {et~al.}(2016)\citenamefont {Hwang},
  \citenamefont {Riggleman},\ and\ \citenamefont {Crocker}}]{hwang16}%
  \BibitemOpen
  \bibfield  {author} {\bibinfo {author} {\bibfnamefont {H.~J.}\ \bibnamefont
  {Hwang}}, \bibinfo {author} {\bibfnamefont {R.~A.}\ \bibnamefont
  {Riggleman}}, \ and\ \bibinfo {author} {\bibfnamefont {J.~C.}\ \bibnamefont
  {Crocker}},\ }\bibfield  {title} {\enquote {\bibinfo {title} {Understanding
  soft glassy materials using an energy landscape approach},}\ }\href@noop {}
  {\bibfield  {journal} {\bibinfo  {journal} {Nature Mat.}\ }\textbf {\bibinfo
  {volume} {15}},\ \bibinfo {pages} {1031} (\bibinfo {year}
  {2016})}\BibitemShut {NoStop}%
\bibitem [{\citenamefont {Morse}\ and\ \citenamefont {Corwin}(2014)}]{morse14}%
  \BibitemOpen
  \bibfield  {author} {\bibinfo {author} {\bibfnamefont {P.~K.}\ \bibnamefont
  {Morse}}\ and\ \bibinfo {author} {\bibfnamefont {E.~I.}\ \bibnamefont
  {Corwin}},\ }\bibfield  {title} {\enquote {\bibinfo {title} {Geometric
  signatures of jamming in the mechanical vacuum},}\ }\href@noop {} {\bibfield
  {journal} {\bibinfo  {journal} {Phys. Rev. Lett.}\ }\textbf {\bibinfo
  {volume} {112}},\ \bibinfo {pages} {115701} (\bibinfo {year}
  {2014})}\BibitemShut {NoStop}%
\bibitem [{\citenamefont {Morse}\ and\ \citenamefont {Corwin}(2017)}]{morse17}%
  \BibitemOpen
  \bibfield  {author} {\bibinfo {author} {\bibfnamefont {P.~K.}\ \bibnamefont
  {Morse}}\ and\ \bibinfo {author} {\bibfnamefont {E.~I.}\ \bibnamefont
  {Corwin}},\ }\bibfield  {title} {\enquote {\bibinfo {title} {Echoes of the
  glass transition in a thermal soft spheres},}\ }\href@noop {} {\bibfield
  {journal} {\bibinfo  {journal} {Phys. Rev. Lett.}\ }\textbf {\bibinfo
  {volume} {119}},\ \bibinfo {pages} {118003} (\bibinfo {year}
  {2017})}\BibitemShut {NoStop}%
\bibitem [{\citenamefont {Kob}\ and\ \citenamefont {Barrat}(1997)}]{kob97b}%
  \BibitemOpen
  \bibfield  {author} {\bibinfo {author} {\bibfnamefont {W.}~\bibnamefont
  {Kob}}\ and\ \bibinfo {author} {\bibfnamefont {{J.-L.}}\ \bibnamefont
  {Barrat}},\ }\bibfield  {title} {\enquote {\bibinfo {title} {Aging effects in
  a Lennard-Jones glass},}\ }\href@noop {} {\bibfield  {journal} {\bibinfo
  {journal} {Phys. Rev. Lett.}\ }\textbf {\bibinfo {volume} {78}},\ \bibinfo
  {pages} {4581} (\bibinfo {year} {1997})}\BibitemShut {NoStop}%
\bibitem [{\citenamefont {{Mendoza-M{\'e}ndez}}\ \emph
  {et~al.}(2022)\citenamefont {{Mendoza-M{\'e}ndez}}, \citenamefont
  {{Peredo-Ortiz}}, \citenamefont {{L{\"a}zaro-L{\'a}zaro}}, \citenamefont
  {{Ch{\'a}vez-Paez}}, \citenamefont {{Ruiz-Estrada}}, \citenamefont
  {{Pacheco-V{\'a}zquez}}, \citenamefont {{Medina-Noyola}},\ and\ \citenamefont
  {{Elizondo-Aguilera}}}]{mendoza22}%
  \BibitemOpen
  \bibfield  {author} {\bibinfo {author} {\bibfnamefont {P.}~\bibnamefont
  {{Mendoza-M{\'e}ndez}}}, \bibinfo {author} {\bibfnamefont {R.}~\bibnamefont
  {{Peredo-Ortiz}}}, \bibinfo {author} {\bibfnamefont {E.}~\bibnamefont
  {{L{\"a}zaro-L{\'a}zaro}}}, \bibinfo {author} {\bibfnamefont
  {M.}~\bibnamefont {{Ch{\'a}vez-Paez}}}, \bibinfo {author} {\bibfnamefont
  {H.}~\bibnamefont {{Ruiz-Estrada}}}, \bibinfo {author} {\bibfnamefont
  {F.}~\bibnamefont {{Pacheco-V{\'a}zquez}}}, \bibinfo {author} {\bibfnamefont
  {M.}~\bibnamefont {{Medina-Noyola}}}, \ and\ \bibinfo {author} {\bibfnamefont
  {L.~F.}\ \bibnamefont {{Elizondo-Aguilera}}},\ }\bibfield  {title} {\enquote
  {\bibinfo {title} {Structural relaxation, dynamical arrest, and aging in
  soft-sphere liquids},}\ }\href@noop {} {\bibfield  {journal} {\bibinfo
  {journal} {J. Chem. Phys.}\ }\textbf {\bibinfo {volume} {157}},\ \bibinfo
  {pages} {244504} (\bibinfo {year} {2022})}\BibitemShut {NoStop}%
\bibitem [{\citenamefont {Silbert}\ \emph {et~al.}(2005)\citenamefont
  {Silbert}, \citenamefont {Liu},\ and\ \citenamefont {Nagel}}]{silbert05}%
  \BibitemOpen
  \bibfield  {author} {\bibinfo {author} {\bibfnamefont {L.~E.}\ \bibnamefont
  {Silbert}}, \bibinfo {author} {\bibfnamefont {A.~J.}\ \bibnamefont {Liu}}, \
  and\ \bibinfo {author} {\bibfnamefont {S.~R.}\ \bibnamefont {Nagel}},\
  }\bibfield  {title} {\enquote {\bibinfo {title} {Vibrations and diverging
  length scales near the unjamming transition},}\ }\href@noop {} {\bibfield
  {journal} {\bibinfo  {journal} {Phys. Rev. Lett.}\ }\textbf {\bibinfo
  {volume} {95}},\ \bibinfo {pages} {098301} (\bibinfo {year}
  {2005})}\BibitemShut {NoStop}%
\bibitem [{\citenamefont {Ikeda}\ \emph {et~al.}(2013)\citenamefont {Ikeda},
  \citenamefont {Berthier},\ and\ \citenamefont {Biroli}}]{ikeda13}%
  \BibitemOpen
  \bibfield  {author} {\bibinfo {author} {\bibfnamefont {A.}~\bibnamefont
  {Ikeda}}, \bibinfo {author} {\bibfnamefont {L.}~\bibnamefont {Berthier}}, \
  and\ \bibinfo {author} {\bibfnamefont {G.}~\bibnamefont {Biroli}},\
  }\bibfield  {title} {\enquote {\bibinfo {title} {Dynamic criticality at the
  jamming transition},}\ }\href@noop {} {\bibfield  {journal} {\bibinfo
  {journal} {J. Chem. Phys.}\ }\textbf {\bibinfo {volume} {138}},\ \bibinfo
  {pages} {12A507} (\bibinfo {year} {2013})}\BibitemShut {NoStop}%
\bibitem [{foo({\natexlab{f}})}]{footsurov}%
  \BibitemOpen
  \bibinfo {note} {A closely related effect was reported in Ref.\
  \cite{surovtsev03}, which showed that the ``boson peak'' frequency
  $\omega_{\rm bp}$ defined by the maximum of $D(\omega)/\omega$ in
  small-molecule glasses \cite{sokolov92} increased with increasing annealing
  time.}\BibitemShut {Stop}%
\bibitem [{\citenamefont {Surovtsev}\ \emph {et~al.}(2003)\citenamefont
  {Surovtsev}, \citenamefont {Shebanin},\ and\ \citenamefont
  {Ramos}}]{surovtsev03}%
  \BibitemOpen
  \bibfield  {author} {\bibinfo {author} {\bibfnamefont {N.~V.}\ \bibnamefont
  {Surovtsev}}, \bibinfo {author} {\bibfnamefont {A.~P.}\ \bibnamefont
  {Shebanin}}, \ and\ \bibinfo {author} {\bibfnamefont {M.~A.}\ \bibnamefont
  {Ramos}},\ }\bibfield  {title} {\enquote {\bibinfo {title} {Density of states
  and light-vibration coupling coefficient in $\rm{B}_2\rm{O}_3$ glasses with
  different thermal history},}\ }\href@noop {} {\bibfield  {journal} {\bibinfo
  {journal} {Phys. Rev. B}\ }\textbf {\bibinfo {volume} {67}},\ \bibinfo
  {pages} {024203} (\bibinfo {year} {2003})}\BibitemShut {NoStop}%
\bibitem [{\citenamefont {Sokolov}\ \emph {et~al.}(1992)\citenamefont
  {Sokolov}, \citenamefont {Kisliuk}, \citenamefont {Soltwisch},\ and\
  \citenamefont {Quitmann}}]{sokolov92}%
  \BibitemOpen
  \bibfield  {author} {\bibinfo {author} {\bibfnamefont {A.~P.}\ \bibnamefont
  {Sokolov}}, \bibinfo {author} {\bibfnamefont {A.}~\bibnamefont {Kisliuk}},
  \bibinfo {author} {\bibfnamefont {M.}~\bibnamefont {Soltwisch}}, \ and\
  \bibinfo {author} {\bibfnamefont {D.}~\bibnamefont {Quitmann}},\ }\bibfield
  {title} {\enquote {\bibinfo {title} {Medium-range order in glasses:
  Comparison of raman and diffraction measurements},}\ }\href@noop {}
  {\bibfield  {journal} {\bibinfo  {journal} {Phys. Rev. Lett.}\ }\textbf
  {\bibinfo {volume} {69}},\ \bibinfo {pages} {1540} (\bibinfo {year}
  {1992})}\BibitemShut {NoStop}%
\end{thebibliography}

%

\end{document}